\documentclass[preprint*]{JHEP3} 

\usepackage{epsfig,multicol}

\newcommand\fverb{\setbox\pippobox=\hbox\bgroup\verb}
\newcommand\fverbdo{\egroup\medskip\noindent%
                        \fbox{\unhbox\pippobox}\ }
\newcommand\fverbit{\egroup\item[\fbox{\unhbox\pippobox}]}
\newbox\pippobox

\title{
Equivalence between Kaluza Klein modes of gravitinos and
goldstinos in brane induced supersymmetry breaking}

\author{Stefania De Curtis\\
I.N.F.N.,
Sezione di Firenze,  I-50019 Sesto F., Italy\\
E-mail: \email{decurtis@fi.infn.it}}

\author{Daniele Dominici\\
Dipartimento di Fisica, Universit\`a di Firenze,  I-50019 Sesto
F., Italy\\
I.N.F.N.,
Sezione di Firenze,  I-50019 Sesto F., Italy\\
E-mail: \email{dominici@fi.infn.it}}

\author{Jos\'e R. Pelaez\\
Dept. de F\'{\i}sica Te\'orica II,
  Univ. Complutense, 28040 Madrid, Spain\\
  E-mail: \email{jrpelaez@fis.ucm.es}}

\received{}             
\revised{}
\accepted{}             

\preprint{\hepth{******}}  

\abstract{We identify the goldstino fields that give mass to the
Kaluza Klein modes of five dimensional supergravity, when
supersymmetry breaking is induced by brane effects. We then proof
the four dimensional Equivalence Theorem that, in renormalizable
gauges, allows for the replacement of Kaluza Klein modes of
helicity $\pm1/2$ gravitinos in terms of goldstinos. Finally we
identify the five dimensional renormalizable gauge fixing that
leads to the Equivalence Theorem.}


\begin{document}
\newcommand{\NP}[1]{ Nucl.\ Phys.\ {#1}}
\newcommand{\ZP}[1]{ Z.\ Phys.\ {#1}}
\newcommand{\RMP}[1]{ Rev.\ of Mod.\ Phys.\ {#1}}
\newcommand{\PL}[1]{ Phys.\ Lett.\ { #1}}
\newcommand{\NC}[1]{ Nuovo Cimento {#1}}
\newcommand{\AN}[1]{ Ann. Phys. {#1}}
\newcommand{\PRep}[1]{Phys.\ Rep.\ {#1}}
\newcommand{\PR}[1]{Phys.\ Rev.\ { #1}}
\newcommand{\PRL}[1]{ Phys.\ Rev.\ Lett.\ { #1}}
\newcommand{\MPL}[1]{ Mod.\ Phys.\ Lett.\ {#1}}
\newcommand{\IJmp}[1]{{\em Int.\ J.\ Mod.\ Phys.\ }{\bf #1}}
\newcommand{\Od}{O}
\newcommand{\qfint}{\int \frac{d^4 q}{(2\pi)^4}}
\newcommand{\qtint}{\int \frac{d^3 \vec{q}}{(2\pi)^3}}
\newcommand{\qmixv}{(q_0,\vec{q},\tau)}
\newcommand{\qmixw}{(q_0,\omega_q,\tau)}
\newcommand{\tr}{\mbox{tr}}
\newcommand{\Tr}{\mbox{Tr}}
\newcommand{\Dim}{\mbox{dim}}
\newcommand{\im}{\mbox{Im}\,}
\newcommand{\re}{\mbox{Re}\,}
\newcommand{\sgn}{\mbox{sgn}}
\newcommand{\diag}{\mbox{diag}}
\newcommand{\hpisqt}{h_\pi^2 (t)}
\newcommand{\fpitsq}{f_\pi^2 (t)}
\newcommand{\ftsq}{f^2(t)}
\newcommand{\fzerosq}{f^2(0)}
\newcommand{\fdot}{\dot f(t)}
\newcommand{\fddot}{\ddot f(t)}
\newcommand{\tti}{\tilde t}
\newcommand{\lagf}{{\cal L}^{(4)}}
\newcommand{\intc}{\int_C dt \int d^3 \vec{x}}
\newcommand{\intcc}{\int_C \! d^4x}
\newcommand{\vxt}{(\vec{x},t)}
\newcommand{\bea}{\begin{eqnarray}}
\newcommand{\eea}{\end{eqnarray}}
\newcommand{\be}{\begin{equation}}
\newcommand{\ee}{\end{equation}}
\def\bc{\begin{center}}
\def\ec{\end{center}}
\newcommand{\ar}{\arrowvert}
\newcommand{\ra}{\rangle}
\newcommand{\la}{\langle}
\newcommand{\da}{^\dagger}
\newcommand{\ov}{\overline}
\newcommand{\cd}{\! \cdot \!}
\newcommand{\tad}{F_\beta}
\newcommand{\f}{\frac}
\newcommand{\dd}{\displaystyle}
\newcommand{\lr}[1]{{ \left( \, #1 \, \right) }}
\newcommand{\gs}{g'_5 \, \! ^2}
\newcommand{\s}{\sigma}
\newcommand{\sbar}{\bar\sigma}
\newcommand{\smu}{\sigma^\mu}
\newcommand{\sbarmu}{\bar\sigma^\mu}

\newcommand{\nn}{\nonumber}
\def\f{\frac}

\def\dd{\displaystyle}
\def\nn{\nonumber}
\def\sgn{{\rm sgn}}
\def\ad{\dot{\alpha}}
\def\ov{\overline}
\def\cP{{P}}\def\vl{{v_\lambda}}
\def\vl{{v_\lambda}}
\def\v{{v}}
\def\bPsi{{\bf \Psi}}
\def\hbPsi{{\bf \hat\Psi}}
\def\hbPsim{{{\bf \hat \Psi^M}}}

\newcommand{\derbar}{\not{\!\partial}}


\section{Introduction}

Extra dimensional models provide a geometrical approach to tackle
the hierarchy problem in particle physics. In addition, they are
suggested by recent developments of string theory. In many of
these models supersymmetry plays a relevant role in generating or
stabilizing the hierarchies. This is for instance the case of the
M-theory extensions of the heterotic   $E_8\times E_8$ string,
which lead to a scenario with two branes at the end of a finite
dimension \cite{Horava:1996ma}. The bulk is populated by
supergravity whereas matter and  gauge fields are constrained to
the branes. For our purposes, we will neglect the six dimensions
compactified on the Calabi-Yau manifold and we will consider an
effective five dimensional  (5D) supergravity with two four
dimensional (4D) walls containing the matter and the gauge
fields. In these models  the supersymmetry breaking can be
induced by brane localized dynamics
\cite{Horava:1996vs,Meissner:1999ja,Bagger:2001ep} leading to
massive gravitinos
 through the super Higgs mechanism.

From the four dimensional point of view, however, the gravitino
becomes a tower of massive Kaluza Klein (KK) gravitinos, whose mass
is due to the brane supersymmetry breaking effects
and the five dimensional compactification.
Actually, the gravitino mass eigenfields
are a complicated combination of the Fourier modes of the 5D
gravitino fields. The zero KK mode was identified in \cite{Meissner:1999ja}
and the whole tower in
\cite{Bagger:2001ep}. Intuitively, gravitinos become massive
because they acquire helicity $\pm1/2$ components
by eating the goldstino  degrees of freedom.
Indeed there is an Equivalence Theorem (ET) for 4D supergravity,
that relates at high energy  the scattering amplitudes of helicity $\pm1/2$
gravitinos to amplitudes where these gravitinos are
replaced by their associated goldstinos \cite{Casalbuoni:1988qd}. This theorem
is very useful in calculating gravitino observables
since it is much easier to handle spin 1/2 particles, as goldstinos,
than complicated helicity projections of Rarita-Schwinger spin 3/2 fields.
One of the most typical applications is the use of the ET
to calculate the cosmological production and decay of massive gravitinos \cite{Kallosh:1999jj},
that impose severe constraints in the parameter space
of supergravity and/or cosmology if standard nucleosynthesis is to be preserved.
It has been recently pointed out that the gravitino Kaluza Klein modes
are likely to pose a bigger challenge to nucleosynthesis \cite{Mazumdar:2000gj}, but
the calculation of their abundance was postponed until the formal approach for
their study  was available. In this work we provide such a
framework. Another common application of the ET is the study of 
perturbative unitarity limits,
since it is well known that tree level amplitudes containing
helicity $\pm1/2$ gravitinos can violate tree level unitarity \cite{Bhattacharya:1988nk}.
In this respect, it has been recently shown that the introduction of
extra dimensions could worsen dramatically the problem of unitarity 
unless the interactions are conveniently suppressed to 
balance the increase of KK states with the available energy \cite{DeCurtis:2003zt}.
In principle, such effects also occur for 5D gravitinos.

The zero mode goldstino for the effective 5D supergravity was
already identified in \cite{Meissner:1999ja,Bagger:2001ep}.
However, the formulation of the ET in extra dimensional models
presents the additional complication that goldstinos are also a
mixture of KK components that depends non-trivially on both the
brane and the bulk dynamics \cite{DeCurtis:2002nd}. Also, the ET
is formulated in the so called renormalizable t'Hooft-Veltman, or
$R_\xi$ gauges, while these kind of models are usually implemented
 in the unitary gauge
(i.e. explicitly removing the goldstinos) \cite{Meissner:1999ja,Bagger:2001ep}.

In this work, we present a complete proof of the ET in
brane induced supersymmetry breaking scenarios, relating
each KK mode of the gravitinos with their corresponding
combination of KK goldstinos. In section 2 we introduce
the details of the model. In section 3 we perform the diagonalization of
 the gravitino and goldstino mass terms
as well as their mixing term. In section 4 we prove the ET, by
identifying first the gravitino and goldstino mass eigenfields
and the four dimensional $R_\xi$ gauge fixing. At the end of the
section we present the gauge fixing in its five dimensional form,
and we study the limit of small extra dimensions where the ET is
particularly simple even in terms of the initial fields. In
section 5 we summarize.

\section{The bulk and brane actions}

The action  that  reproduces all the main
features of gaugino condensation \cite{Horava:1996vs}
 in M-theory \cite{Horava:1996ma}
at the level of the five dimensional effective theory
is given in \cite{Bagger:2001ep}.
The fifth dimension is compactified on the $S^1/Z_2$ orbifold, obtained
by identifying $x^5\leftrightarrow -x^5$. Customarily $x^5$ varies in the interval $[0,\pi \kappa]$.
Introducing a 1/2 factor to avoid overcounting, the bulk action is defined to be:
\begin{equation}
  \label{eq:sbulk}
  S_{bulk}=\frac {1}{2}\int d^4 x\int_{-\pi\kappa}^{+\pi\kappa}d\,x^5 {\cal L}_{bulk}
\end{equation}
where ${\cal L}_{bulk}$ is the 5D supergravity Lagrangian density
\cite{cremmer}. The five dimensional coordinates are
$x^M=(x^m,x^5)$ and the fields are fluctuations over the following
background:
\begin{equation}
  < g_{MN}>=\left(
  \begin{array}{cc}
\eta_{mn}& 0\\
0&r^2
  \end{array}
\right)
\end{equation}
where $R=r/M_5$ is the physical compactification radius,
 $M_5= \kappa^{-1}$ being
the 5D (reduced) Planck mass  related to the 4D
 Planck mass by $M_4^2=\pi R M_5^3$.
Two gravitino fields are needed to form a generalized Majorana 5D
spinor, in particular, for the gravity multiplets, we have:
$\Psi=(\psi^1_\alpha,\overline{\psi^2}^{\dot \alpha})$ and
$\overline\Psi=(\psi^{2\,\alpha},\overline{\psi^1}_{\dot
\alpha})$. We assign even $Z_2$-parity to
$e_m^a,e_{5\hat5},\psi^1_m,\psi^2_5$ and odd $Z_2$-parity to
$e_5^a,e_{m\hat5},\psi^2_m,\psi^1_5$, here $e_M^A$ is the
f\"unfbein and $\hat 5$ is the fifth tangent-space index. We will
denote by $e_4$ ($e_5$) the determinant of the vierbein
(f\"unfbein), although, since we will be interested in fermion
bilinears only, we will set it to 1 for simplicity in most cases.
 From the 4D point of view, the
zero modes of $e_m^a$  and $\psi^1_m$, with spins (2,3/2)
respectively, form a massless $N=1$ gravitational supermultiplet.
The fields $e_{5\hat 5}$
and $\psi_5^2$ form, together with the $B_5$ component 
of the graviphoton $B_M$, a $N=1$ massless chiral multiplet.
In addition one has  an infinite Kaluza Klein (KK) series of
multiplets of N=2 supergravity with spins (2,3/2,3/2,1) and masses
$M_n^2=n^2/R^2$, $n=1,2...$. As we will see in detail below, the
KK gravitino tower gets mass through an infinite series of super
Higgs effects  by eating the KK modes of the goldstino field.

Brane physics is not relevant for our purposes. We will follow
\cite{Bagger:2001ep} and imagine that the branes 
are located at orbifold fixed points without tension-like terms and
without generating a warp factor so that the bosonic
vacuum is flat. We also assume
that fields living
in the brane are
integrated out leaving a constant superpotential vacuum
expectation value (vev) on each brane. 
It has been shown in \cite{Bagger:2001ep} that these
vev's can spontaneously break the remaining N=1 supersymmetry.
Since the physics can be different on both branes the vev's are
parametrized by two  constants $P_0$ and $P_\pi$ (which we have
taken real for simplicity) with dimension $mass^3$. Thus
 \cite{Horava:1996vs,Meissner:1999ja,Bagger:2001ep}:
\begin{equation}
  \label{eq:sbrane}
  S_{brane}=\frac{\kappa^2}{2}\int d^4x\int_{-\pi\kappa}^{\pi\kappa} dx^5
[\delta(x^5) P_0+\delta(x^5-\pi\kappa)
P_\pi]\psi^1_m\sigma^{mn}\psi_n^1+{\rm h.c.}
\end{equation}

In order  to study the super Higgs effect and the equivalence between
KK goldstinos and gravitinos,
 it is enough to consider the 5D fermionic bilinear terms:
\bea S_{2f}^{(5)} & = & {1 \over 2} \, \int d^4 x \int_{- \pi
\kappa}^{+ \pi \kappa} d x^5 \, \left\{ {r \over \kappa}
\epsilon^{mnpq} \left( \ov{\psi^1}_m \ov{\sigma}_n \partial_p
\psi_q^1 - \psi_m^2 \sigma_n \partial_p \ov{\psi^2}_q \right)
\right. \nn \\ && + {2 \over \kappa} e_4 \left( - \psi_m^2
\sigma^{mn} \partial_5 \psi_n^1 + \ov{\psi^1}_m \ov{\sigma}^{mn}
\partial_5 \ov{\psi^2}_n + \psi_m^2 \sigma^{mn} \partial_n
\psi_5^1 \right. \nn \\ & & \left. -  \ov{\psi^1}_m
\ov{\sigma}^{mn} \partial_n \ov{\psi^2}_5 + \psi_5^2 \sigma^{mn}
\partial_m \psi_n^1 - \ov{\psi^1}_5 \ov{\sigma}^{mn} \partial_m
\ov{\psi^2}_n \right) \nn \\ && + \left. \left(  e_4 \kappa^2
\left[ \delta(x^5) {P_0} + \delta(x^5 - \pi \kappa) P_{\pi}
\right] \psi_m^1 \sigma^{mn} \psi_n^1 + {\rm \; h.c.} \right)
\right\} \, . \label{s52f} \eea

In contrast to \cite{Bagger:2001ep}, where the $\Psi^5$ field is
eliminated by going to the unitary gauge, we are interested in
identifying the goldstinos eaten by the gravitinos to acquire
their mass through the Higgs mechanism. For that reason it is
convenient to perform several transformations that will allow us
to diagonalize simultaneously the gravitino and goldstino mass
matrices as well as their mixing term. Let us follow
\cite{Dolan:1984fm} and transform the fields
\begin{eqnarray}
  \label{eq:dolan}
\Psi_m\rightarrow\Psi_m+\frac{1}{\sqrt{6}} \Gamma_m\Gamma^5\Psi_5, \qquad
\Psi_5\rightarrow \frac{2 r }{\sqrt{6}}\Psi_5
\end{eqnarray}
(an equivalent transformation, but with slightly different
redefinitions, can be  found also in \cite{Meissner:1999ja}). We
obtain: \bea \label{postdolan} S_{2f}^{(5)} & = & {1 \over 2} \,
\int d^4 x \int_{- \pi \kappa}^{+ \pi \kappa} d x^5 \, \left\{ {r
\over \kappa} \epsilon^{mnpq} \left( \ov{\psi^1}_m \ov{\sigma}_n
\partial_p \psi_q^1 - \psi_m^2 \sigma_n \partial_p \ov{\psi^2}_q
\right) \right.
\\ &&
-i\frac{r
e_4}{\kappa}\left(\psi_5^2\s^m\partial_m\ov{\psi^2}_5+\ov{\psi^1}_5
\sbar^m\partial_m\psi^1_5 \right) +\frac{2
e_4}{\kappa}\left(\ov{\psi^1}_5\partial_5\ov{\psi^2}_5-\psi^2_5
\partial_5\psi^1_5
\right)\nn\\
&&+{i\,\sqrt{6}e_4\over 2\kappa}
\left(\ov{\psi^1}_5\sbar^m\partial_5\psi^1_m+\psi^2_m\s^m\partial_5\ov{\psi^2}_5+{\rm
\; h.c.} \right) \nn\\&&+ {2 e_4 \over \kappa} \left( - \psi_m^2
\sigma^{mn} \partial_5 \psi_n^1 + \ov{\psi^1}_m \ov{\sigma}^{mn}
\partial_5 \ov{\psi^2}_n
\right) \nn \\ && + \left. \left(  e_4 \kappa^2  \left[
\delta(x^5) {P_0} + \delta(x^5 - \pi \kappa) {P_{\pi}}  \right]
\psi_m^1 \sigma^{mn} \psi_n^1 + {\rm \; h.c.} \right) \right\} \,
\nn \\ && + \left. \left(  e_4 \kappa^2  \left[ \delta(x^5) {P_0}
+ \delta(x^5 - \pi \kappa) {P_{\pi}}  \right]
\left[{-i\sqrt{6}\over 2}\left(\psi^1_m\s^m\ov{\psi^2}_5
\right)+\ov{\psi^2}_5\ov{\psi^2}_5\right]+ {\rm \; h.c.} \right)
\right\} \, \nn \eea Note that, in this way, we have generated a
standard kinetic term for the $\Psi_5$ field that was not present
before. We will need this term to prove the equivalence between
helicity $\pm 1/2$ gravitinos and goldstinos. We can now write the
4D reduction of the above Lagrangian, recalling that the even
fields $\psi^{\rm even}=\psi^1_m, \psi^2_5$ have the following
Fourier expansion:
\begin{equation}
\psi^{\rm even}(x^5) =  {1 \over \sqrt{\pi r}} \left[ \psi^{\rm
even}_{0} + \sqrt{2} \sum_{\rho=1}^{\infty} \psi^{\rm even}_{\rho}
\cos(\rho M_5 x^5) \right] \, , \label{eq:5to4even}
\end{equation}
whereas the odd fields  $\psi^{\rm odd}=\psi^2_m, \psi^1_5$ have:
\begin{equation}
\psi^{\rm odd} (x^5)={\sqrt{2} \over \sqrt{\pi r}}
 \sum_{\rho=1}^{\infty} \psi^{\rm odd}_{\rho} \sin(\rho M_5 x^5)\, ,
\label{eq:5to4odd}
\end{equation}
consistently with their $Z_2$-parity assignments. After
integration of the $x^5$ coordinate, we find the following 4D
Lagrangian (we consider only the part of the Lagrangian which is
quadratic in the fermion fields):
\bea {\cal L}_{2f}^{(4)} & = & \left\{ {1 \over 2}
\epsilon^{mnpq}\left(\ov{\psi^1}_{p,0} {\ov{\sigma}}_q
\partial_m \psi^1_{n,0}+\sum_{\rho=1}^{\infty}
\ov{\psi^1}_{p,\rho} {\ov{\sigma}}_q \partial_m \psi^1_{n,\rho}
+\sum_{\rho=1}^{\infty} \ov{\psi^2}_{p,\rho} {\ov{\sigma}}_q
\partial_m \psi^2_{n,\rho} \right) \right. \nn \\
& - & {i \over 2} \left( \ov{\psi^2}_{5,0} \sbar^m \partial_m
\psi^2_{5,0} + \sum_{\rho=1}^{\infty} \ov{\psi^2}_{5,\rho} \sbar^m
\partial_m \psi^2_{5,\rho} + \sum_{\rho=1}^{\infty}
\ov{\psi^1}_{5,\rho}  \sbar^m \partial_m \psi^1_{5,\rho} \right)
\nn \\
& + & {2 \over r} \sum_{\rho=1}^{\infty} \left( \rho
M_5 \right) \left( \psi^2_{m,\rho} \sigma^{mn} \psi^1_{n,\rho}
-\psi^1_{5,\rho}  \psi^2_{5,\rho}
\right)
\nn \\
& - & {i \sqrt{6}  \over 2 r} \sum_{\rho=1}^{\infty} \left( \rho
M_5 \right) \left( \ov{\psi^1}_{5,\rho} \sbar^m \psi^1_{m,\rho}
+\psi^2_{5,\rho} \s^m \ov{\psi^2}_{5,\rho} \right)
\nn \\
& + & {\kappa^2 \over 2 \pi r}  \, {P}_0 \left[ \left(
\psi^1_{m,0} + \sqrt{2} \sum_{\rho=1}^{\infty}
\psi^1_{m,\rho} \right) \sigma^{mn} \left(
\psi^1_{n,0} + \sqrt{2} \sum_{\sigma=1}^{\infty}
\psi^1_{n,\sigma} \right) \right. \\
& + & \left. \left( \ov{\psi^2}_{5,0} + \sqrt{2}
\sum_{\rho=1}^{\infty} \ov{\psi^2}_{5,\rho} \right)  \left(
\ov{\psi^2}_{5,0} + \sqrt{2} \sum_{\sigma=1}^{\infty}
\ov{\psi^2}_{5,\sigma} \right) \right]\nn  \\
& + &  {\kappa^2 \over 2 \pi r}  \, \left[{P}_{\pi}
\left( \psi^1_{m,0} + \sqrt{2} \sum_{\rho=1}^{\infty}
(-1)^{\rho} \psi^1_{m,\rho} \right) \sigma^{mn} \left(
\psi^1_{n,0} + \sqrt{2} \sum_{\sigma=1}^{\infty}
(-1)^{\sigma} \psi^1_{n,\sigma} \right) \right.
\nn \\
& + & \left. \left( \ov{\psi^2}_{5,0} + \sqrt{2}
\sum_{\rho=1}^{\infty} (-1)^\rho \ov{\psi^2}_{5,\rho} \right)
\left( \ov{\psi^2}_{5,0} + \sqrt{2}
\sum_{\sigma=1}^{\infty}(-1)^\sigma
\ov{\psi^2}_{5,\sigma} \right) \right]\nn  \\
& - & i {\sqrt{6} \kappa^2 \over 4 \pi r}  \, {P}_0 \left(
\psi^1_{m,0} + \sqrt{2} \sum_{\rho=1}^{\infty} \psi^1_{m,\rho}
\right) \sigma^{m} \left( \ov{\psi^1}_{5,0} + \sqrt{2}
\sum_{\sigma=1}^{\infty}
\ov{\psi^2}_{5,\sigma} \right) \nn  \\
& - & \left. i {\sqrt{6} \kappa^2 \over 4 \pi r}  \, {P}_{\pi}
\left(\psi^1_{m,0} + \sqrt{2} \sum_{\rho=1}^{\infty} (-1)^{\rho}
\psi^1_{m,\rho} \right) \sigma^{m} \left( \ov{\psi^2}_{5,0} +
\sqrt{2} \sum_{\sigma=1}^{\infty} (-1)^{\sigma}
\ov{\psi^2}_{5,\sigma} \right) \right\}
+   {\rm \; h.c.}\nn
\label{l42f}
\eea

\section{Mass matrix eigenstates}
The gravitino mass matrix has been shown to diagonalize under the
following transformations. First, by defining \cite{Bagger:2001ep}
\be
\psi_{m,\rho}^{\pm} = {\psi_{m,\rho}^1 \pm \psi_{m,\rho}^2 \over
\sqrt{2}} \, , \; (\rho > 0) \, , \;\;\;\;\;\; P_{\pm} =
{\kappa^3 \over 2 \pi} \left( {P}_0 \pm {P}_{\pi} \right) \, ,
\label{plusminusfields} \ee 
in the basis
${\bf\Psi}_m=(\psi^1_{m,0},\psi^+_{m,1},\psi^-_{m,1},\psi^+_{m,2},
\psi^-_{m,2},...)^T$ the mass matrix reads:
\be
{\cal M}_{3/2} = {1 \over R}
\left(
\begin{array}{c|cc|cc|c}
\cP_+ & \cP_- & \cP_- & \cP_+ & \cP_+ & \ldots \\
\hline
\cP_- & \cP_+ + 1 & \cP_+ & \cP_- & \cP_- & \ldots \\
\cP_- & \cP_+ & \cP_+ - 1 & \cP_-       & \cP_- & \ldots \\
\hline
\cP_+ & \cP_-  & \cP_- & \cP_+ + 2 & \cP_+ & \ldots \\
\cP_+ & \cP_-     & \cP_- & \cP_+ & \cP_+ - 2 & \ldots \\
\hline
\ldots & \ldots & \ldots & \ldots & \ldots & \ldots \\
\end{array}
\right) \,.
\label{3/2matrix}
\ee
We are also interested in the goldstino mass terms and in the
mixing terms between goldstinos and gravitinos. Note that
the greatest simplification is achieved if we define the goldstino
basis in a slightly different way than was done for gravitinos in eq.(\ref{plusminusfields}),
namely, by defining \be \psi_{5,\rho}^{\pm} = {\mp\psi_{5,\rho}^1 +
\psi_{5,\rho}^2 \over \sqrt{2}} \, , \; (\rho > 0) \, .
\label{plusminusfields5} \ee 
In the ${\bf
\Psi}_5=(\psi^2_{5,0},\psi^+_{5,1},\psi^-_{5,1},\psi^+_{5,2},\psi^-_{5,2},...)^T$
basis, we find that  the goldstino  mass matrix is nothing but the
gravitino mass matrix just given above in eq.(\ref{3/2matrix}).
Thus the 4D Lagrangian density suffers a dramatic simplification:
\begin{eqnarray}
  \label{eq:Lsimple}
  {\cal L}^{(4)}_{2f}&=&\frac{1}{2}\epsilon^{mnpq}\ov{\bf \Psi}_m^T\sbar_n\partial_p{\bf \Psi}_q
-\frac{i}{2}\ov{\bf \Psi}_5^T\sbar^m\partial_m{\bf \Psi}_5\nn\\
&+&
{\bf \Psi}_m^T{\cal M}_{3/2}\s^{mn}{\bf \Psi}_n+
\ov{\bf \Psi}_5^T{\cal M}_{3/2}{\bf \Psi}_5
- i \sqrt{\frac{3}{2}}{\bf \Psi}_m^T{\cal M}_{3/2}\s^{m}\ov{\bf \Psi}_5+{\rm h.c.}
\end{eqnarray}
Of course, for calculations we are interested in the mass eigenstates, which
are obtained as follows.
As shown in ref.\cite{Bagger:2001ep}, 
the eigenvalues $\lambda$ of the matrix  $R {\cal M}_{3/2}$
satisfy the following equation:
\be \tan(\pi\lambda) = \frac {4\pi P_+}{\pi^2 (P_-^2 -P_+^2)+4} ,
\label{eigenvalues} \ee 
and therefore the mass eingenvalues are
given by 
\be m^a_{3/2}\equiv m^{(\pm\rho)}_{3/2}=\frac 1 R \left
\{ \frac 1 \pi \arctan \left [  \frac {4\pi P_+ }{\pi^2 (P_-^2
-P_+^2)+4} \right ]\pm\rho\right \}. \ee 
Here $a$ enumerates the
components of the vectors ${\bf\Psi}_m$ and ${\bf\Psi}_5$ while
$\rho=0, 1, 2,\dots$ refers to the KK modes. The orthogonal matrix $Q$
that diagonalizes the gravitino mass matrix 
\be {\cal
M}_{3/2}^D=Q^T {\cal M}_{3/2} Q=\diag \{
m^{(0)}_{3/2},m^{(1)}_{3/2},m^{(-1)}_{3/2},\cdots \} \ee has the
following form
\be
Q =
\left(
\begin{array}{cccccc}
\rule[-3mm]{.0mm}{8mm}
\dd{q_0} & \dd{q_{1}}& \dd{-q_{-1}} & \dd {q_{2}} &
\dd{-q_{-2}} & \ldots \\
\rule[-3mm]{.0mm}{10mm}\dd\frac {q_0 c_0} {(\lambda_{0}-1)} &\dd\frac {q_1 c_1} {(\lambda_{1}-1)}
&\dd\frac {-q_{-1}c_{-1}} {(\lambda_{-1}-1)}  &\dd\frac {q_2c_2} {(\lambda_{2}-1)}
& \dd\frac {-q_{-2}c_{-2}} {(\lambda_{-2}-1)}& \ldots \\
\rule[-3mm]{.0mm}{10mm}
\dd\frac {q_0c_0} {(\lambda_{0}+1)} &\dd\frac {q_1c_1} {(\lambda_{1}+1)}   &
\dd\frac {-q_{-1}c_{-1}} {(\lambda_{-1}+1)} & \dd  \frac {q_2c_2} {(\lambda_{2}+1)}
&\dd \frac {-q_{-2}c_{-2}} {(\lambda_{-2}+1)}  & \ldots \\
\rule[-3mm]{.0mm}{10mm}\dd\frac {q_0\lambda_{0}} {(\lambda_{0}-2)} & \dd\frac {q_1\lambda_{1}} {(\lambda_{1}-2)} &
\dd \frac {-q_{-1}\lambda_{-1}} {(\lambda_{-1}-2)} &
\dd \frac {q_2\lambda_{2}} {(\lambda_{2}-2)} & \dd \frac {-q_{-2}\lambda_{-2}}
{(\lambda_{-2}-2)} & \ldots \\
\rule[-3mm]{.0mm}{10mm} \dd\frac {q_0\lambda_{0}} {(\lambda_{0}+2)}&
\dd\frac {q_1\lambda_{1}} {(\lambda_{1}+2)}
& \dd \frac {-q_{-1}\lambda_{-1}} {(\lambda_{-1}+2)} &
\dd \frac {q_2\lambda_{2}} {(\lambda_{2}+2)} &\dd \frac {-q_{-2}\lambda_{-2}}
{(\lambda_{-2}+2)} & \ldots \\
\ldots & \ldots & \ldots & \ldots & \ldots & \ddots \\
\end{array}
\right) \,,
\label{Qmatrix}
\ee
with 
\be q_{\rho}= \left [1- 2c_\rho ^2 (\lambda_{\rho} \frac
{\partial \Sigma_O}{\partial \lambda_{\rho}} +\Sigma_O
)-2\lambda_{\rho}^2 ( \lambda_{\rho} \frac {\partial
\Sigma_E}{\partial \lambda_{\rho}} +\Sigma_E )\right ]^{-1/2}, 
\ee
\be 
c_\rho= \frac {1} {P_-}\left [ (P_-^2 -P_+^2)( 2
\lambda_{\rho}^2 \Sigma_E +1) +P_+\lambda_{\rho} \right ] ,
\ee
where 
\be \Sigma_E(\lambda)=\Sigma_{\rho\, even}\frac {1}{
\lambda^2-\rho ^2}= -\frac {1}{2 \lambda^2}+\frac {\pi} {4
\lambda} \left [ \frac {1+\cos (\lambda\pi)}{\sin
(\lambda\pi)}\right ], \ee
 and 
\be
\Sigma_O(\lambda)=\Sigma_{\rho\, odd}\frac {1}{ \lambda^2-\rho ^2}
=-\frac {\pi} {4 \lambda} \left [ \frac {1-\cos
(\lambda\pi)}{\sin (\lambda\pi)}\right ].
\ee 
Therefore we
introduce the mass eigenvector spinors
${\bf\hat\Psi}_m=(\hat\psi^1_{m,0},\hat\psi^+_{m,1},\hat\psi^-_{m,1},\hat\psi^+_{m,2},
\hat\psi^-_{m,2},...)^T$ and ${\bf
\hat\Psi}_5=(\hat\psi^2_{5,0},\hat\psi^+_{5,1},\hat\psi^-_{5,1}
,\hat\psi^+_{5,2},\hat\psi^-_{5,2},...)^T$ related to $\bPsi_m$
and $\bPsi_5$ by 
\be \hbPsi_m=Q^T \bPsi_m, \quad \hbPsi_5=Q^T
\bPsi_5 \label{eigspin}, 
\ee 
in terms of which the Lagrangian
given in eq.(\ref{eq:Lsimple}) can be re-expressed as
\begin{eqnarray}
  \label{eq:Lmoresimple}
  {\cal L}^{(4)}_{2f}&=&\frac{1}{2}\epsilon^{mnpq}\ov{\hbPsi}_m^T\sbar_n\partial_p{\hbPsi}_q
-\frac{i}{2}\ov{\hbPsi}_5^T\sbar^m\partial_m{\hbPsi}_5\nn\\
&+&
{\hbPsi}_m^T{\cal M}_{3/2}^D\s^{mn}{\hbPsi}_n+
\ov{\hbPsi}_5^T{\cal M}_{3/2}^D{\hbPsi}_5
- i \sqrt{\frac{3}{2}}{\hbPsi}_m^T{\cal M}_{3/2}^D\s^{m}\ov{\hbPsi}_5+{\rm h.c.}
\end{eqnarray}
In our discussion we have assumed $P_0\neq P_\pi$; when $P_0=- P_\pi$
the transformation in eq.(\ref{Qmatrix}) is singular. 
In that case a $N=1$ SUSY is left
unbroken \cite{Horava:1996ma}
 and a field redefinition to get the fermion bilinear
lagrangian in a diagonalized form has been identified \cite{Meissner:1999ja}.

Once we have identified the  mass and 
mixing terms of the goldstino and gravitino
KK tower, we turn to the proof of the ET.

\section{The Equivalence Theorem}

The Equivalence Theorem we will prove here is an extension of the
corresponding theorem in spontaneously broken supergravity in four
dimensions, first suggested in \cite{Fayet:vd} and proved in
\cite{Casalbuoni:1988qd}: at center of mass energies $E>>m_{3/2}$,
the S-matrix elements with external helicity $\pm 1/2$ gravitinos
(longitudinal gravitinos) are equivalent to the S-matrix elements
with corresponding goldstinos. In order to use the known results
it is convenient to rewrite the Lagrangian density in eq.
(\ref{eq:Lmoresimple}) using  Majorana spinors defined in terms
of the corresponding Weyl ones as \be \psi^{(M)}=\left(
  \begin{array}{c}
\psi_\alpha\\
{\ov\psi}^{\ad}
 \end{array}
\right ),\quad
{\ov\psi}^{(M)}=\left(
\psi^\alpha ,
{\ov\psi}_{\ad}
\right ).
\ee
The result is
\begin{eqnarray}
  \label{eq:Lmajorana}
  {\cal L}^{(4)}_{2f}&=&\frac{1}{2}\epsilon^{mnpq}
\ov{ \hat {\bf\Psi}}{}^{(M)\,T}_m\gamma_5\gamma_n\partial_p \hat
{\bf\Psi}{}^{(M)}_q
-\frac{i}{2}\ov{ \hat {\bf\Psi}}{}^{(M)\,T}_5\gamma^m\partial_m \hat {\bf\Psi}{}^{(M)}_5\\
&+&
 \hat {\bf\Psi}{}^{(M)\,T}_m {\cal M}_{3/2}^D\gamma^{mn}\hat {\bf\Psi}{}^{(M)}_n+
\ov{ \hat {\bf\Psi}}{}^{(M)\,T}_5{\cal M}_{3/2}^D {\bf
{\hat\Psi}}_5{}^{(M)} - i \sqrt{\frac{3}{2}} \ov{\bf
{\hat\Psi}}{}{}^{(M)\,T}_m {\cal M}_{3/2}^D\gamma^{m}{ \hat
{\bf\Psi}{}^{(M)}_5}, \nn
\end{eqnarray}
with $\gamma^{mn}$ given in Appendix A. Apart from the mass,
${\bf\Psi}_5$ and metric sign conventions, we find that  ${\cal
L}^{(4)}_{2f}$ is nothing but a sum of infinite copies
 of the Lagrangian (2.6) in
ref.\cite{Casalbuoni:1988qd}, one for each component of the
gravitino and goldstino vectors. This Lagrangian was the starting
point for the proof of the ET between $\pm$1/2 helicity
gravitinos and goldstinos. 

\subsection{$R_\xi$ gauge fixing}
The basic idea for the proof was to
use the analogous of $R_\xi$ gauges in the non abelian gauge
theories. Contrary to the unitary gauges used in
\cite{Meissner:1999ja,Bagger:2001ep}, within these so called
renormalizable or t'Hooft-Veltman gauges, the gravitinos and the
goldstinos coexist in the Lagrangian, and is possible to relate
them at high energy through the ET. We recall that the main
virtue of these generalized $R_\xi$ gauges is to cancel the
mixing term between the gravitino and the goldstino fields. This
is achieved by adding to the Lagrangian density a gauge fixing
term. It is easy to check that, in the present case, the gauge
fixing term  has the following form:
 \be {\cal L}^{(4)}_{GF}=-\frac {i}{
2} \xi \sum_{a}{\ov{\hat F_a}} \derbar \hat F_a \label{FdF}, \ee 
with 
\be
\hat F_a=\gamma^m  { \hat{\Psi}}{}^{(M)}_{m,a}
 -\sqrt{\frac 3 2} { m}_{3/2}^{a}\frac {1 }{\xi\derbar}{
 \hat{\Psi}}{}^{(M)}_{5,a}.
\ee 
It is interesting to rewrite the gauge fixing term in a 5D form.
In fact, since we have a copy
of the Lagrangian for each component of the vectors
${\bf\hat\Psi}_m=(\hat\psi^1_{m,0},\hat\psi^+_{m,1},\hat\psi^-_{m,1},\hat\psi^+_{m,2},
\hat\psi^-_{m,2},...)^T$ and ${\bf
\hat\Psi}_5=(\hat\psi^2_{5,0},\hat\psi^+_{5,1},\hat
\psi^-_{5,1},\hat\psi^+_{5,2},\hat\psi^-_{5,2},...)^T$ (see eq.
(\ref{eq:Lmoresimple})), we  need a gauge fixing term for each
component, each one relating a gravitino KK state  with its
corresponding KK goldstino mode. As we have already remarked,
these gauge fixing terms are found by requiring the cancellation
of the gravitino-goldstino mixing term in eq.
(\ref{eq:Lmajorana}). The 4D gauge fixing Lagrangian can be
written:
\begin{equation}
  \label{eq:GFtower}
  {\cal L}^{(4)}_{GF}=-\frac {i} {2} \,\xi \,{\overline {\bf \hat F}{}^T} \derbar {\bf \hat F}; \quad
{\bf \hat F}=\gamma^m  {\bf  \hat{\Psi}}^{(M)}_m -\sqrt{\frac 3 2}
{\cal  M}^D_{3/2}\frac {1 }{\xi\derbar}{ \bf \hat{\Psi}}^{(M)}_5.
\end{equation}

We show in Appendix B how to generate the 4D gauge fixing term in
eq.(\ref{FdF}) from the original 5D Lagrangian in eq.
(\ref{postdolan}). We check there that the 5D gauge fixing term
can be written as
\begin{eqnarray}
  \label{eq:5DGF}
 {\cal L}^{(5)}_{GF}=
-\frac {i}{ 2} \,\xi \, {\ov H} \Gamma^m\partial_m H=
 -\frac {i}{ 4} \,\xi \,\left[ \overline {H^1} \sbar^m\partial_m H^1
+\overline {H^2} \sbar^m\partial_m H^2 \right]+ {\rm h.c.};
\end{eqnarray}
where $\Gamma^m$ is given in Appendix A, $H$ is a generalized
Majorana 5D spinor: $H=(H^1_\alpha,\overline{H^2}^{\dot \alpha})$
and $\overline H =(H^{2\,\alpha},\overline{H^1}_{\dot \alpha})$,
with
\begin{eqnarray}
  \label{eq:5DGFfunctions}
H^1&=&\s^m \ov{\psi^1}_m-\sqrt{\frac 3 2} \frac {1
}{\xi\partial^2}\partial_m\s^m
\left(\partial_5\ov{\psi^1}_5-\kappa^3  \left[ \delta(x^5)
{P_0} + \delta(x^5 - \pi \kappa) {P_{\pi}}  \right]\ov{\psi^2}_5\right)\\
H^2&=&\s^m \ov{\psi^2}_m-\sqrt{\frac 3 2} \frac {1
}{\xi\partial^2}\partial_m\s^m\partial_5\ov{\psi^2}_5
\end{eqnarray}
We have checked that the gauge fixing term needed
in 5D is precisely the one that cancels the 5D mixing terms in eq.(\ref{postdolan}). Note
that it is made of two gauge fixing functions, since we have two
gravitinos, exactly as it happens when we have several gauge
fields and we have to gauge fix
each one of them. Also, note that
the gauge fixing term breaks the 5D Lorentz invariance:
 first because the mixings induced by the branes
 break it  by the $\delta$-terms, and second because
of the presence of $\partial^m$ derivatives. This is also what
happens, for instance, in electrodynamics when using Coulomb
gauges, which are originally thought to deal with static
problems, where three coordinates are enough to describe an
otherwise four dimensional theory.

\subsection{The Equivalence Theorem for KK modes}

The sum in eq.(\ref{FdF}) is over the components of the
vectors ${ \hat{\bf\Psi}}{}^{(M)}_{m}$ and
${\hat{\bf\Psi}}{}^{(M)}_{5}$ and ${ m}_{3/2}^{a}$ are the
eigenvalues of the gravitino mass matrix. From the path integral
point of view, a gauge fixing as in eq.(\ref{FdF}) is a delta
functional $\delta(\derbar F)$, which intuitively corresponds to
imposing ${\derbar F}=0$ throughout the calculations. That is, we
get a relation between $\derbar \gamma^m
{\hat{\Psi}}{}^{(M)}_{m,a} $ and ${ \hat{\Psi}}{}^{(M)}_{5,a}$ for
each component $a$. This is precisely what we need to relate the
on-shell amplitudes of helicity $\pm$1/2
 gravitinos and their KK excitations with the
corresponding amplitudes for goldstinos.

The rigorous proof \cite{Casalbuoni:1988qd} makes use of  the BRS
invariance to get a set of relevant Ward identities  leading to
the corresponding relations between S-matrix elements for
external gravitinos and goldstinos. The main steps are the
following. In the high energy regime, that is, for energies much
larger than the gravitino mass ($s>>m_{3/2}^2$), one can
approximate the wave function corresponding to the longitudinal
components of the gravitino  as \be \psi^{\pm 1/2}=-\bar E^{\pm
1/2}_m\psi^m \ee where \be
 E^{\pm 1/2}_m (p)\sim \left(
i \sqrt{\frac{2}{3}}
\frac{p_m}{m_{3/2}}+{\cal{O}}\left(\frac{m_{3/2}}{p}\right)\right)u^{\pm1/2}
(p) \ee with $u^{\pm1/2} (p)$ a Dirac spinor with appropriate
polarization. Using the BRS invariance and the relation between
the goldstino and the gravitino wave operators one obtains
\begin{equation}
  \label{eq:ETgeneric}
  S(A,\psi^{\pm1/2},B)=S(A,\psi_5,B)
+{\cal{O}}(m_{3/2}/\sqrt{s})
\end{equation}
where $S(A,\psi^{\pm1/2}(\psi_5),B)$ denotes the $S$ matrix elements
for the  longitudinal gravitino
(goldstino) and any other $A$ and $B$ physical states.

We can rephrase all the proof in terms of the components of the
vectors ${ \hat{\bf\Psi}}{}^{(M)}_{m}$ and
${\hat{\bf\Psi}}{}^{(M)}_{5}$, and formulate an Equivalence
Theorem relating the $S$ matrix elements involving the
longitudinal components of  KK gravitinos  with the ones involving
the corresponding KK goldstinos. At energies where we can neglect
the masses of the gravitinos involved in the scattering amplitude
under consideration, it reads:
\begin{equation}
  \label{eq:ETKK}
  T(\hat \Psi^{\pm 1/2}_{a},\hat \Psi^{\pm1/2}_{b},...\hat \Psi^{\pm1/2}_{c})
= T(\hat \Psi^{(M)}_{5\,a},\hat \Psi^{(M)}_{5\,b},...\hat
\Psi^{(M)}_{5\,c})+O({\rm max}[\vert m^a_{3/2}\vert/\sqrt{s},\vert m^b_{3/2}\vert/\sqrt{s},...])
\end{equation}
where $\hat \Psi^{\pm1/2}_{a}$ denotes the $a$ component of the
vector $\hat {\bf\Psi}^{\pm 1/2}=-\bar E^{\pm 1/2}_m{\hat {\bf
\Psi}}^{(M)\,m}$ and $\hat \Psi^{(M)}_{5\,a}$ denotes the $a$
component of the vector  $\hat {\bf\Psi}_5^{(M)}$.  Note that the
ET holds for the mass eigenfields and for energies higher than
the mass of the heaviest KK mode in the amplitude. Finally, let
us remark that the ET is most useful in the Landau gauge,
$\xi\rightarrow\infty$, where the  gauge dependent complicated
goldstino kinetic terms coming from the gauge fixing function
cancel, and we just have to deal with the usual simple
propagators for spin 1/2 fields.

\subsection{The $R\rightarrow0$ limit}

It is instructive to obtain the values of the $Q$ elements
in the $R\sim \kappa^{1/3}\rightarrow0$ limit, ($R$ and $\kappa$ are related
by requiring a finite 4D Planck mass $M_4$). In such case
 the physics is basically four dimensional with a very tiny fifth dimension,
which is a natural limit in the brane induced supersymmetry breaking
scenarios. Indeed the scale associated to the orbifold
 $(\pi \kappa)^{-1}$  comes
out to be of the order 10$^{15}$ GeV, while the scale of the
supersymmetry breaking $[(P_0+P_\pi)/2]^{1/3}$ is of the order
 10$^{13}$ GeV (see for instance \cite{Nilles:2001ap}
and references therein).
 In this case $P_+<<1$ and  the lightest gravitino mass
turns out to be \be m_{3/2}^{(0)}\sim \frac {P_+}{R}\sim \frac 1
2 \frac {P_0+P_\pi}{M_4^2} \sim 200~{\rm GeV}. \ee 
In this limit,
the matrix   $Q$ reduces to: \be \label{k3} Q = \left(
\begin{array}{cccc}
\rule[-3mm]{.0mm}{8mm}
\dd 1+O(\kappa^6) & \dd{O(\kappa^3)}& \dd{O(\kappa^3)} & \ldots \\
\rule[-3mm]{.0mm}{8mm}
\dd {O(\kappa^3)} & \dd 1+O(\kappa^6)& \dd{O(\kappa^3)} & \ldots \\
\rule[-3mm]{.0mm}{8mm}
\dd {O(\kappa^3)} & \dd{O(\kappa^3)} &\dd 1+O(\kappa^6)&  \ldots \\
\ldots & \ldots & \ldots & \ddots \\
\end{array}
\right) \,, \label{Qmatrixapprox} \ee 
That is, the mass
eigenfields are basically those initially in the Lagrangian except
for an $O(\kappa^3)$ correction. Consequently, 
\bea
  \label{eq:ETKK1}
T(\hat \Psi^{\pm 1/2}_{a},\hat \Psi^{\pm1/2}_{b},...\hat
\Psi^{\pm1/2}_{c}) &\simeq T( \Psi^{(M)}_{5\,a}, \Psi^{(M)}_{5\,b},...
\Psi^{(M)}_{5\,c})
\eea
In other words, in this limit  we can calculate
scattering amplitudes involving helicity $\pm1/2$ gravitinos in a
simpler way by using directly the vertices of the interaction
Lagrangian for goldstinos, without the need for a rotation.
Note that terms we are neglecting, with respect to the complete
formulation of the ET in eq.(\ref{eq:ETKK}) are {\it suppressed} 
by $O(\kappa^3)$ factors (to obtain the precise order in $\kappa$ we would
need to know the superpotential).
Of course, to get higher order corrections in $\kappa$
we need the complete formulation.


\section{Conclusions}

We have presented a proof of the Equivalence Theorem
for massive Kaluza Klein gravitino modes in 5D supergravity with brane induced supersymmetry
breaking. This theorem holds
for energies higher than the mass of any of the gravitinos involved,
and allows to
perform calculations of amplitudes containing on-shell helicity $\pm1/2$ gravitinos
by substituting them with their corresponding goldstino fields.

In particular, we have
identified the four and five dimensional gauge fixing functions
from which  the ET follows.
We have also
identified the goldstino field combinations
that correspond to each gravitino Kaluza Klein mode, providing
the expressions of the rotation matrices  to obtain these eigenfields
in the general case. Finally, we have studied the behavior of these
transformations in the limit of a small extra dimension.

The results presented here can be useful in further studies \cite{Antonioyyo}
of observables involving massive Kaluza Klein gravitinos,
for instance in a cosmological context, that could provide bounds
on the parameters of these supersymmetry breaking models.

\section*{Acknowledgments}
The authors thank A. L. Maroto for his comments and careful
reading of the manuscript,
as well as
partial  support
from the INFN-CICYT Florence-Madrid Collaboration.
J.R.P. thanks the hospitality of the INFN Sezione di Firenze
as well as partial
support from the Spanish CICYT projects
PB98-0782 and BFM2000 1326.

\begin{appendix}

\section{Notation}
We follow the conventions of  Wess and Bagger, \cite{Wess:cp}, in particular we recall that
\begin{eqnarray}
&&  \eta_{mn}=(-1,1,1,1), \quad
\Gamma^5=\left(
  \begin{array}{cc}
-i &0\\
0&i
  \end{array}
\right)= i\,\gamma^5, \quad
\Gamma^m=\left(
  \begin{array}{cc}
0 &\sigma^m\\
\bar\sigma^m&0
  \end{array}
\right), \\
&&\epsilon_{0123}=-1,  \quad \sigma^m=(-I,\vec\sigma),
  \quad \bar\sigma^m=(-I,-\vec\sigma),\\
&&
\sigma^{mn}=\frac 1 4 (\sigma^m \bar\sigma^n-\sigma^n \bar\sigma^m), \quad
\bar\sigma^{mn}=\frac 1 4 ( \bar\sigma^m \sigma^n- \bar\sigma^n \sigma^m),
\end{eqnarray}
\begin{eqnarray}
\gamma^{mn}=\left(
  \begin{array}{cc}
\sigma^{mn} &0\\
0&{\ov\sigma^{mn}}
  \end{array}
\right).
\end{eqnarray}

\section{Gauge fixing term in 5D supergravity}

Let us show   how to generate the 4D gauge fixing term in eq.
 (\ref{eq:GFtower}) from the original 5D Lagrangian in eq.(\ref{postdolan}). First, we
rewrite  it in terms of Weyl spinors. For that purpose it is
convenient to note that
\begin{equation}
  \label{eq:isthesame}
  {\bf { \overline{\hat  F}}}{}^T \derbar {\bf \hat F}= ({\overline {-\gamma_5 \bf \hat F}}){}^T
\derbar (-\gamma_5 {\bf \hat F})
\end{equation}
where $-\gamma_5 {\bf \hat F}$ is a Majorana spinor (whereas ${\bf \hat F}$ is not).
Then we can simply re-express the 4D gauge fixing Lagrangian
 in terms of one Weyl spinor ${\bf \hat f}$
as follows:
\begin{eqnarray}
  \label{eq:GFWeyl4D}
 &&-\gamma_5{\bf \hat F}=\left(
 \begin{array}{c}
{\bf \hat f}_\alpha\\
{\bf \bar {\hat f}}{}^{\dot\alpha}
 \end{array}\right),\\
&&{\bf \hat f}=\s^m  \ov{\bf  \hat{\Psi}}_m -\sqrt{\frac 3 2}
{\cal  M}^D_{3/2}\frac {1 }{\xi\partial^2}\partial_m\s^m\ov{ \bf
\hat{\Psi}}_5
,\\
&&   {\cal L}^{(4)}_{GF}=-\frac {i}{ 2} \,\xi \,{\overline {\bf
\hat f}}{}^T \sbar^m\partial_m {\bf \hat f}+ {\rm h.c.}
\end{eqnarray}
Let us now recall that the transformations
$\hbPsi_m=Q^T \bPsi_m, \hbPsi_5=Q^T \bPsi_5 $ are orthogonal, so that
the above gauge fixing term can be recast by replacing
$\hbPsi_m\rightarrow \bPsi_m, \hbPsi_5\rightarrow \bPsi_5 $ and
${\cal M}^D_{3/2}\rightarrow{\cal M}_{3/2}=Q {\cal M}^D_{3/2} Q^T$.
Recalling that the $\psi^\pm$ fields are related to the $\psi^{1,2}$ fields
by eqs.(\ref{plusminusfields}), (\ref{plusminusfields5}), we can write
\begin{eqnarray}
  \label{eq:pmto12}
{\bf \Phi_m}=
\left(
  \begin{array}{c}
\psi^1_{m,0}\\
\psi^1_{m,1}\\
\psi^2_{m,1}\\
\psi^1_{m,2}\\
\psi^2_{m,2}\\
\vdots
\end{array}
\right)=\Omega {\bf \Psi_m}\equiv
\left(
  \begin{array}{cccccc}
1& & & & & \\
 &1 \over \sqrt{2} & 1 \over \sqrt{2}& & &\\
 &1 \over \sqrt{2} & -1 \over \sqrt{2}& & &\\
&&&1 \over \sqrt{2} & 1 \over \sqrt{2}& \\
&& &1 \over \sqrt{2} & -1 \over \sqrt{2}& \\
&&&&&\ddots
\end{array}
\right)
\left(
  \begin{array}{c}
\psi^1_{m,0}\\
\psi^+_{m,1}\\
\psi^-_{m,1}\\
\psi^+_{m,2}\\
\psi^-_{m,2}\\
\vdots
\end{array}
\right)
\end{eqnarray}
and
\begin{eqnarray}
  \label{eq:pmto125}
{\bf \Phi}_5= \left(
  \begin{array}{c}
\psi^2_{5,0}\\
\psi^1_{5,1}\\
\psi^2_{5,1}\\
\psi^1_{5,2}\\
\psi^2_{5,2}\\
\vdots
\end{array}
\right)=\Omega_5 {\bf \Psi}_5\equiv \left(
  \begin{array}{cccccc}
1& & & & & \\
 &-1 \over \sqrt{2} & 1 \over \sqrt{2}& & &\\
 &1 \over \sqrt{2} & 1 \over \sqrt{2}& & &\\
&&&-1 \over \sqrt{2} & 1 \over \sqrt{2}& \\
&& &1 \over \sqrt{2} & 1 \over \sqrt{2}& \\
&&&&&\ddots
\end{array}
\right)
\left(
  \begin{array}{c}
\psi^2_{5,0}\\
\psi^+_{5,1}\\
\psi^-_{5,1}\\
\psi^+_{5,2}\\
\psi^-_{5,2}\\
\vdots
\end{array}
\right).
\end{eqnarray}
Since $\Omega=\Omega^T=\Omega^{-1}$, $\Omega_5=\Omega_5^T=\Omega_5^{-1}$,
again we can simply rewrite the gauge fixing term
as
\begin{eqnarray}
  \label{eq:GFWeyl12}
   {\cal L}^{(4)}_{GF}=\frac {-i}{ 2} \,\xi \,{\overline {\bf h}}{}^T \sbar^m\partial_m {\bf h}+ {\rm h.c.}; \quad
{\bf h}=\s^m  \ov{\bf  \Phi}_m -\sqrt{\frac 3 2} \frac {1
}{\xi\partial^2}\partial_m\s^m \Omega{\cal  M}_{3/2}\Omega_5\ov{
\bf \Phi}_5.
\end{eqnarray}
Let us write explicitly
\begin{eqnarray}
  \label{eq:brbu}
  \Omega{\cal M}_{3/2}\Omega_5 \ov{ \bf \Phi}_5 =
\left(
  \begin{array}{c}
P_+\bar\psi^2_{5,0}+\sqrt{2}\sum_{\rho=1}^\infty( P_0+(-1)^\rho  P_\pi)\bar\psi^2_{5,\rho}\\
\sqrt{2}\left[P_-\bar\psi^2_{5,0}+\sqrt{2}\sum_{\rho=1}^\infty( P_0+(-1)^{\rho+1}  P_\pi)\bar\psi^2_{5,\rho}\right]\\
0\\
\sqrt{2}\left[P_+\bar\psi^2_{5,0}+\sqrt{2}\sum_{\rho=1}^\infty( P_0+(-1)^{\rho+2}  P_\pi)\bar\psi^2_{5,\rho}\right]\\
0\\
\vdots
  \end{array}
\right)+
\left(
  \begin{array}{c}
0\\
-\bar\psi^1_{5,1}\\
\bar\psi^2_{5,1}\\
-2\bar\psi^1_{5,2}\\
2\bar\psi^2_{5,2}\\
\vdots
  \end{array}
\right)
\end{eqnarray}
The first vector is related to the physics on the branes and the second to the bulk.
It is straightforward to check that the gauge fixing term in eq.(\ref{eq:GFWeyl12}) comes
from a 5D gauge fixing term:

\begin{eqnarray}
  \label{eq:5DGFa}
 {\cal L}^{(5)}_{GF}=
-\frac {i}{ 2} \,\xi \, {\ov H} \Gamma^m\partial_m H=
 -\frac {i}{ 4} \,\xi \,\left[ \overline {H^1} \sbar^m\partial_m H^1
+\overline {H^2} \sbar^m\partial_m H^2 \right]+ {\rm h.c.};
\end{eqnarray}
where $\Gamma^m$ is given in Appendix A, $H$ is a generalized
Majorana 5D spinor: $H=(H^1_\alpha,\overline{H^2}^{\dot \alpha})$
and $\overline H =(H^{2\,\alpha},\overline{H^1}_{\dot \alpha})$,
with
\begin{eqnarray}
  \label{eq:5DGFfunctionsa}
H^1&=&\s^m \ov{\psi^1}_m-\sqrt{\frac 3 2} \frac {1
}{\xi\partial^2}\partial_m\s^m
\left(\partial_5\ov{\psi^1}_5-\kappa^3  \left[ \delta(x^5)
{P_0} + \delta(x^5 - \pi \kappa) {P_{\pi}}  \right]\ov{\psi^2}_5\right),\\
H^2&=&\s^m \ov{\psi^2}_m-\sqrt{\frac 3 2} \frac {1
}{\xi\partial^2}\partial_m\s^m\partial_5\ov{\psi^2}_5.
\end{eqnarray}

 It is easy to verify that this term exactly cancels the mixing terms between the
 goldstinos and the gravitinos
in the 5D Lagrangian density given in eq.(\ref{postdolan}).

\end{appendix}

\end{document}